\RequirePackage[l2tabu,orthodox]{nag}
\documentclass[entropy,article,pdftex,accept,oneauthor,preprints]{Definitions/mdpi}

\firstpage{1}
\pubvolume{xx}
\issuenum{x}
\articlenumber{x}
\pubyear{2022}
\copyrightyear{2022}
\history{Received: date; Accepted: date; Published: date}

\usepackage{color,xspace}
\usepackage{Definitions/macros_gpi}

\usepackage[utf8]{inputenc}
\usepackage{float}
\usepackage{textcomp}
\usepackage{graphicx}
\ifx\hypersetup\undefined
  \AtBeginDocument{%
    \hypersetup{unicode=true}
  }
\else
  \hypersetup{unicode=true}
\fi

\usepackage{natbib}
\usepackage{amstext}
\usepackage{siunitx}
\usepackage[colorinlistoftodos,prependcaption]{todonotes}
\usepackage{cleveref}

\usepackage{babel}
\usepackage[linesnumbered,lined,tworuled,commentsnumbered]{algorithm2e}
\SetKwFor{Loop}{Loop:}{}{}

\Title{Snowballing Nested Sampling}
\newcommand{\orcidauthorA}{0000-0003-0426-6634} 
\Author{Johannes Buchner $^{1}$ orcid number:\orcidauthorA}
\AuthorNames{0000-0003-0426-6634}
\address{%
$^1$ \quad Max Planck Institute for Extraterrestrial Physics, Giessenbachstrasse, 85748 Garching, Germany
}
\corres{Johannes Buchner}
\firstnote{Submitted to International Workshop on Bayesian Inference and Maximum Entropy Methods in Science and Engineering, XXXX, 2023.} 

\abstract{
A new way to run nested sampling, combined with realistic MCMC proposals to generate new live points, is presented. Nested sampling is run with a fixed number of MCMC steps.
Subsequently, snowballing nested sampling extends the run to more and more live points. 
This stabilizes the MCMC proposal of later MCMC proposals,
and leads to pleasant properties, including 
that the number of live points and number of MCMC steps do not have to be calibrated,
that the evidence and posterior approximation improves as more compute is added and can be diagnosed with convergence diagnostics from the MCMC community.
Snowballing nested sampling converges to a ``perfect'' nested sampling run with infinite number of MCMC steps.
}

\begin{document}

\maketitle

\section{Introduction}

Nested sampling \citep{Skilling2004} is a
popular \citep{Ashton2022} Monte Carlo algorithm for Bayesian inference, 
estimating the posterior distribution and the Bayesian evidence. 
A set of $K$ live
points are initially drawn from the prior, and their likelihoods evaluated.
In each iteration of the algorithm, the live point with the lowest
likelihood is discarded and becomes a dead point. 
It is replaced with a new point sampled from
the prior, under the constraint that the likelihood of the discarded
point is exceeded. 
This likelihood-restricted prior
sampling (LRPS) can be achieved in various ways 
(see \citet{Buchner2021} for an extensive review).

\citet{Skilling2004} originally proposed Markov Chain
Monte Carlo (MCMC) random walk algorithms for the LRPS.
This is still the most
popular choice for high-dimensional problems, including slice sampling variants (e.g. \cite{Handley2015}), recently compared for efficiency and reliability in \cite{Buchner2022}.
Such algorithms
start from a randomly chosen live point and make $M$ transitions
to yield the new live point. To avoid a collapse of the live
point set, and sufficiently explore the likelihood-restricted prior,
$M$ has to be chosen large enough. 
However, as \citet{Salomone2018} discusses, mild correlations
between live points are acceptable. 
Thus, a suitably efficient LRPS algorithm has to be
chosen by the user, as well as the size of the live points, $K$,
which regulates the resolution of the sampling and the uncertainty
on the outputs. Since $M$ is not known before-hand, to get reliable outputs
\cite{Higson2019} recommended to repeatedly run nested sampling
with ever-doubling $M$ until the Bayesian evidence stabilizes. However,
this may discard information built with substantial
computational resources.

This work highlights that there is an interplay
between $M$ and $K$: If the number of live points is large, nested
sampling discards only a small fraction of the prior in each iteration
($1/K$ of its probability mass), and a large set of live points is
representing the likelihood distribution of the restricted prior.
The higher $K$, and the slower nested sampling traverses the likelihood,
the more acceptable correlations between points should be.


At termination, nested sampling provides posterior samples from the
dead points. Each sampled point is weighted by the likelihood times
the prior volume discarded at that iteration. Because
nested sampling iterates with a fixed speed through the prior, the
number of samples with non-negligible weight may be low. Two solutions have been proposed for this: (1) to keep intermediate ``phantom'' points \citep{Handley2015} from the MCMC chains to represent the dead point, instead of a single estimate. A drawback here is memory storage. (2) to dynamically widen $K$ in a selected likelihood interval (dynamic nested sampling, \citealp{Higson2017}).
The latter can rerun likelihood intervals where most of the posterior
weight is, bulking the posterior samples. The uncertainty on the evidence is dominated by the stochastic sampling of a limited number  of $K$ live points up to this point. Therefore, for enhanced precision on the evidence, all iterations have to be extended to more live points.


In this work, we present a relatively derivative form of nested sampling which (1) avoids choosing $M$ and $K$ carefully beforehand, (2) does not discard previous computations, (3) achieves arbitrarily large effective number of posterior samples as compute increases, and (4) can be diagnosed for convergence despite an imperfect LRPS.

\section{Method}

Instead of improving nested sampling convergence with
a limited Markov Chain by increasing the number of steps $M$, we
increase the number of live points $K$.

\subsection{Basic algorithm}
The algorithm proposed here, Snowballing Nested Sampling, is:

\begin{algorithm}[H]
\DontPrintSemicolon
$K \leftarrow 1$\;
\Loop{}{
	run Nested Sampling with K live points\;
	$K \leftarrow K + 1$
}
\end{algorithm}

At each iteration, this algorithm produces nested sampling results
(evidence, posterior samples). ``Running nested sampling'' means
running with a MCMC LRPS with a fixed number of steps, $M$, until
termination. 
For example, one can adopt a nested sampling termination
criterion that makes the integral contribution remaining in the live points
negligible (e.g., $Z_{\mathrm{live}}/Z_{i}<10^{-6}$).

\subsection{Convergence analysis}

The rate of compression in (vanilla) nested sampling is $\frac{K-1}{K}$ \citep{Chopin2010}. Thus, after $t$ iterations, the likelihood-restricted prior has volume $\left(\frac{K-1}{K}\right)^t$.
The same restricted prior volume can be reached after $t'=2\times t$ iterations with $K'=1+\sqrt{1-1/K}\approx2$ times the number of live points.
With twice the number of iterations, there are twice as many dead points. Loosely speaking, each dead point from a run with $K$ is split into two dead points from a run with $K'$ live points.
The noise in the nested sampling estimate due to more dead points is also reduced, because the variance is inversely proportional to $K$ \citep[][]{chopin2007contemplating,skilling2009nested}. However, here, we focus on analysing the convergence requirements of the LRPS. 

We assume an MCMC random walk with a fixed number of steps $M$ is adopted as an LRPS method. With the restricted prior compressing twice as slowly, and each dead point being split into two dead points, naturally, we find the following equivalence:
A nested sampling run with $M$ steps and $K$ live points is comparable to a run with $K \times 2$ live points but $M/2$ steps, in terms of correlations between live points that could be problematic for nested sampling.
Based on this, $K\to\infty$ with fixed $M$ achieves the properties of $M\to\infty$ with fixed $K$, i.e., a perfect LRPS algorithm for vanilla nested sampling.
For the latter case, \cite{Chopin2010} proved unbiasedness and convergence of $Z$ and the posterior distribution,. A perfect LRPS algorithm is reached with a proposal kernel that is unbounded (or has the support of the prior), as $M\to\infty$. Therefore, the results of \cite{Chopin2010} also hold for snowballing NS, as $K\to\infty$.

\subsection{Improved algorithm}
For practical purposes, the algorithm above can be improved slightly.
Some LRPS methods need at least a few times the dimensionality of
the parameter to build effective transition kernels. Therefore the
iterations with the lowest $K$ can be skipped, and we start at $K_0$. Since
there may be some computation overhead beyond model evaluations in nested
sampling, $K$ can be incremented in steps of $K_{\mathrm{inc}}$.

It is also not necessary to discard runs from previous iterations and start from scratch. 
Firstly, a subsequent nested sampling run can be initialised with the same initial prior samples as in the previous iterations, and $K_{\mathrm{inc}}$ additional ones. Subsequently in each nested sampling iteration, a live point is discarded, and replaced with a new point. In continuous likelihood functions without plateaus, the call to the LRPS at each dead point, creating a new prior sample based on a threshold $L_\mathrm{min}$, has a uniquely identifiable input, $L_\mathrm{min}$. Therefore, the LRPS call can be memoized. That is, the result of the function is stored with the argument, $L_\mathrm{min}$, as its identifier, and when later the  function evaluation with the same argument is needed, the result is returned based on a look up. Only if the identifier is not stored yet, the LRPS call at that $L_\mathrm{min}$ needs to be made.
See \cite{skilling2006nested, Higson2019, Speagle2020, Buchner2022} for previous work on resuming or combining nested sampling runs.

Put together, the improved algorithm is:

\begin{algorithm}[H]
\DontPrintSemicolon
$K \leftarrow K_0$\;
run Nested Sampling with K0 live points\;
\Loop{}{
	$K \leftarrow K + K_\mathrm{inc}$\;
	resume Nested Sampling with K live points\;
}
\end{algorithm}

\section{Results}

\begin{figure}
\centering
\includegraphics[width=0.5\textwidth]{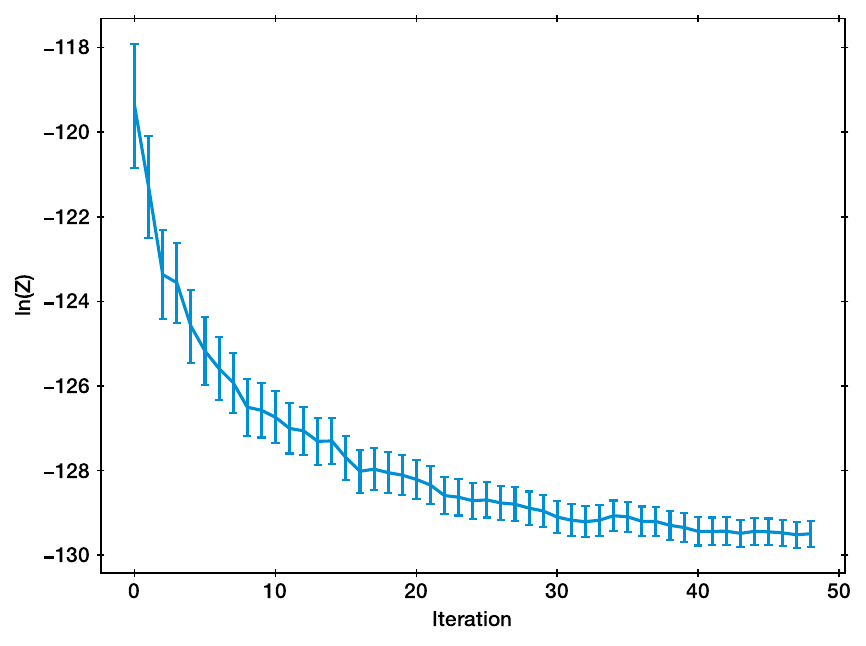}
\caption{\label{fig:rosenZ}Estimate of $\ln(Z)$ for each algorithm iteration.}
\end{figure}

We test the algorithm on the Rosenbrock function over a flat prior
from -10 to +10 in $d=20$ dimensions. The likelihood is:
\[
{\cal \ln L}(\text{\ensuremath{\theta}})=-2\times\sum_{i}\left(100\times\left(\theta_{i+1}-\theta_{i}^{2}\right){}^{2}+\left(1-\theta_{i}\right)^{2}\right)
\]
With $K_{\text{0}}=20$, and $K_{\mathrm{inc}}=20$, and $M=20$,
we observe that $\ln(Z)$ evolves as shown in Figure~\ref{fig:rosenZ}.
From a initial high value, the estimate declines approximately proportional
to $1/(M\times K)$ towards the (presumably true) value if nested
sampling was run with $K=\infty$.

\section{Discussion}

We have presented an iterative form of applying nested sampling. It can
be understood as a kind of dynamic nested sampling. The proposed algorithm
has several amicable properties:
\begin{enumerate}
\item the number of live points and the number of steps does not have to
be chosen carefully,
\item an initial quick look result is available,
\item the evidence estimate improves as more compute is added,
\item in multi-modal settings, additional peaks are discovered as more compute is added,
\item the convergence can be observed and tested with conventional methods based on MCMC,
\item the posterior sample size increases proportionally to the compute,
\item it is easy to implement.
\end{enumerate}

The convergence as more compute is added, the revisiting and further exploration of missed peaks, and the possibility to use conventional MCMC convergence diagnostics, are properties shared with diffusive nested sampling \citep{brewer2011diffusive}. That algorithm is based on likelihood levels. Instead of monotonically switching to higher likelihoods, the sampler can randomly go up or down in levels, and add more samples (live points) there. Unlike diffusive nested sampling, snowballing nested sampling provides an uncertainty estimate on the evidence $Z$.

Finally, we make some remarks on adaptive MCMC algorithms. \citet{Salomone2018} pointed out that on-the-fly adaptation of MCMC algorithms is problematic, and a 
warm-up nested sampling run should be performed, the adapted proposal kernel (parameters) at each iteration stored, and applied without adaptation in a second, fresh nested sampling run.
If the MCMC transition kernel is created based on estimating
a sample covariance matrix of the live point population, 
this estimate stabilizes as $K$ increases.
Newly added $K_\mathrm{inc}$ live points only mildly alter the estimate as $K\to\infty$.
Another popular adaptation is to alter the proposal scale in Metropolis random walks to target a certain acceptance rate.
If the proposal scale is chosen based on the acceptance rate of the previous $K$ points and their proposal scale, through a weighted mean targeting an acceptance rate of 23.4 per cent, then we have a situation of a vanishing adaptation \citep[see][]{andrieu2008tutorial} with convergence to the true posterior.
Furthermore, as the sampling is more faithful, the live points are also more faithfully distributed, which in turn improves the sampling, leading to a positive feedback effect. This accelerates the convergence further.

\acknowledgments{I thank Robert Salomone and Nicolas Chopin for feedback on an early draft, and the two anonymous referees for helpful comments.}

\reftitle{References}
\bibliography{stats}

\end{document}